\documentclass[preprint]{elsarticle}

\usepackage{amsmath}
\usepackage{amssymb}
\usepackage{times}

\begin{document}

\title{Relativistic BCS-BEC crossover of a two-species Fermi gas with number density asymmetry at zero temperature}

\author{Hao Guo}
\ead{guohao@uchicago.edu}
\author{Chih-Chun Chien\corref{cor1}}
\ead{chihchun@uchicago.edu}
\author{Yan He}
\ead{heyan@uchicago.edu}
\cortext[cor1]{Corresponding author}

\address{James Franck Institute and Department of Physics,
 University of Chicago, Chicago, Illinois 60637}

\date{\today}
\begin{abstract}
We systematically study relativistic two-species fermions with tunable attractive interactions and number-density asymmetry at zero temperature. In general, a Bardeen-Cooper-Schrieffer (BCS) - Bose-Einstein Condensation (BEC) - relativistic BEC (RBEC) crossover is observed. A generalized BCS ground state and its stability are analyzed. The homogeneous superfluid phase can become unstable and we consider phase separation in real space for neutral systems. 
In the nonrelativistic limit, our results are consistent with well-known results. In addition to a BCS-BEC crossover similar to that in the nonrelativistic case, in the strongly attractive regime gapless excitations of antifermions and a RBEC state are observed. We address how different phases respond to number density asymmetry and present predictive phase diagrams of BCS-BEC-RBEC crossover. 
\end{abstract}


\maketitle

\section{Introduction}\label{sec:intro}
It is well-known that in nonrelativistic systems, by tuning attractive interactions between fermions, a BCS superfluid of degenerate fermions can be smoothly connected to a Bose-Einstein condensate (BEC) of fermion pairs. For nonrelativistic Fermi gases, it was shown \cite{Leggett,Eagles} that the BCS ground state captures important features in both limits. Study of intermediate phases in between these two limits is called BCS-BEC crossover, which has been intensely studied in high-temperature superconductors \cite{MR93,Sawatzky} and ultra-cold Fermi gases \cite{Jin4,Grimm2,Ketterle3}. In ultra-cold Fermi gases, it was demonstrated that number density asymmetry \cite{ZSSK06,Rice1} or heteronuclear pairing \cite{Wille08} can also be realized.

Recently, the idea of BCS-BEC crossover has been applied to relativistic systems of fermions \cite{boundquark,NishidaPRD05,AbukiNPA05,ZhuangPRD07a,ZhuangPRD07b,WangPRD07}. This work is motivated by two observations. The first comes from color superconductivity of quarks, which has drawn intense attention in high-energy physics \cite{Nardulli,Wilczek00}. At very high baryon density, the ground state of quantum chromodynamics (QCD) is believed to be color superconducting states such as the color-flavor locking (CFL) state or the two-flavor superconducting (2SC) state \cite{Wilczek00}. These states have diquark pairs, which are similar to Cooper pairs of electrons in conventional superconductors. At low baryon density, the system becomes hadronic matters \cite{Baym,BaymPRL}. It was shown \cite{boundquark} that as baryon density decreases, diquark pairs with small pair size comparable to quark-quark distance may emerge. The condensate of these tightly bound diquark pairs are similar to the BEC of Cooper pairs in the BCS-BEC crossover of nonrelativistic fermionic systems. Therefore there may exist a crossover from a color superconducting state with weakly bound quark pairs (of large size) at high baryon density to a (relativistic) BEC state of diquarks at low baryon density before the system becomes hadronic matters \cite{NishidaPRD05,Rezaeian,ZhuangPRD07b}. Perturbative methods, which can only be applied at high baryon densites, are not appropriate for studying this crossover.  

The second observation comes from the fact that a relativistic BEC (RBEC) state may appear even in a nonrelativistic Fermi system if the attractive interaction is strong enough \cite{ZhuangPRD07a}. In an RBEC phase, there is an appreciable density of antifermions.
 Typically, when the $s$-wave scattering length $a$ of a nonrelativistic system approaches the limit $a\sim(\sqrt{2}m)^{-1}$, relativistic effects will become significant since the binding energy of a fermion pair is $\sim\hbar^{2}/(2ma^{2})$. Here $m$ is the fermion mass. In addition to these observations, a crossover in QCD similar to a BCS-BEC crossover of nonrelativistic fermions may also be possible in a pion superfluid or in the interior of compact stars \cite{TDSon01,ZhuangPRD05,ZhuangPRD07c,Shovkovy}.

We study possible phases of a two-species relativistic Fermi gas with presumably tunable attraction. Our approach is inspired by BCS-Leggett theory of nonrelativistic Fermi gases. Number density asymmetry (or polarization), which has been intensely studied in ultra-cold Fermi gases, is included in our calculations. This is important because many interesting phases such as polarized superfluid, phase separation in real space, Larkin-Ovchinnikov-Fulde-Ferrell (LOFF) state \cite{FFLO}, may be observed in nonrelativistic neutral Fermi gases \cite{SR06}. To compare with those results, we only study neutral relativistic Fermi gases.

Past work on relativistic BCS-BEC crossover introduced models including one species of fermions \cite{ZhuangPRD07b} as well as models including two-species of fermions \cite{Chatterjee08,WangPRD07,Brauner08}. Considerations of two (or more) species of fermions are natural since pairing among quarks with different color or flavor indices may exist at moderate baryon densities. Ref.~\cite{Chatterjee08} implemented a variational approach with a coherent state. Missing from this work is an analysis of the stability of superfluid phases. Ref.~\cite{WangPRD07} studied a model with auxiliary boson fields. Conversion between fermions and bosons induces indirect fermion-fermion attraction. Although such a boson-fermion model is possible for ultra-cold fermions across a Feshbach resonance, it may not be necessary for providing a mechanism for diquark pairing. Here we consider effective attraction between fermions without any auxiliary boson field.

We will present phase diagrams as a function of polarization and the strength of the attractive interaction. Following Ref.~\cite{Stability}, we analyze the mechanical stability of homogeneous polarized superfluids. When they are unstable, one expects a phase in which a superfluid and a normal Fermi gas can be found in separated spatial regimes. This is an example of phase separation (or mixed states). This inhomogeneous phase has been discussed in neutral fermionic systems such as ultra-cold fermions \cite{ChienPRL} and quark matter in neutron stars \cite{Shovkovy}. However, phase separation may not be stable for charged fermions \cite{Buballa}. In charged Fermi gases with number density asymetry, there is another instability, magnetic instability (or chromo-magnetic instability in color superconductivity), which is associated with the existence of an imaginary Meissner mass of gauge bosons \cite{HuangPRD04}. In the mechanically unstable regimes of homogeneous superfluids, there are other possible phases in additional to phase separation. For example, the LOFF state \cite{FFLO} is also possible in the weakly attractive regime, but here we do not consider this state since we are mainly interested in the strongly attractive regime. A third possibility are chiral condensates with $\langle\bar{\psi}\psi\rangle\neq 0$, which become dominant when baryon density is low enough. A phase of competition and/or coexistence between (R)BEC states of $\psi\psi$-condensates and $\bar{\psi}\psi$-condensates may exist before the chiral condensate phases completely dominate \cite{Rezaeian,Baym,BaymPRL}. We do not address these more complicated situations here. Collective excitations beyond mean field theory are also interesting and some discussions can be found in \cite{Brauner08}.

This paper is organized as follows. In Sec.~\ref{sec:theo} we present the basic theoretical framework of our mean field theory for two-species Fermi gases with fermion-fermion interactions. We implement both a Nambu-Gorkov formalism and a Bogoliubov canonical transformation and derive equations of state. In Sec.~\ref{sec:stab} we address the important stability criterion by considering properties of the number susceptibility matrix.  Phase separation, which is possible in the unstable regimes of homogeneous superfluids, is also discussed. In Sec.~\ref{sec:result}, we present our numerical results and discuss their implications.

\section{Theoretical Framework}\label{sec:theo}
\subsection{Nambu-Gorkov Formalism}
We consider the following Lagrangian 
\begin{equation}
L=\sum_{\sigma=\uparrow,\downarrow}\bar{\psi}_{\sigma}(i\gamma^{\mu}\partial_{\mu}-m_{\sigma}+\mu_{\sigma}\gamma^{0})\psi_{\sigma}+L_{I},
\end{equation}
where $\psi_{\sigma}$, $\bar{\psi}_{\sigma}$ are Dirac four-component spinors with fermion masses $m_{\sigma}$, and $L_{I}$ denotes the attractive interactions between the two species of fermions in the $J^{P}=0^{+}$ channel \cite{OhsakuPRB02,AlfordPRD03}. The fermions are assumed to be neutral. The metric is $g^{\mu\nu}=\mbox{diag}(1,-1,-1,-1)$. 
We use the spin indices $\sigma=\uparrow,\downarrow$ to denote the two species. In this paper we approximate $m_{\uparrow}=m_{\downarrow}=m$. The interaction $L_{I}$ is of the form
\begin{equation}
L_{I}=-g(\psi^{T}_{\uparrow}Ci\gamma_{5}\psi_{\downarrow})(\bar{\psi}_{\downarrow}i\gamma_{5}C\bar{\psi}^{T}_{\uparrow}),
\end{equation}
where $g$ is the attractive coupling constant and $C=i\gamma^{0}\gamma^{2}$ is the charge conjugation matrix. We introduce the gap function $\Delta=-g\langle\psi^{T}_{\uparrow}Ci\gamma_{5}\psi_{\downarrow}\rangle$, which corresponds to the order parameter. From mean field approximation the Lagrangian becomes
\begin{eqnarray}\label{eq:LMF}
L_{MF}=\sum_{\sigma=\uparrow,\downarrow}\bar{\psi}_{\sigma}(i\gamma^{\mu}\partial_{\mu}-m+\mu_{\sigma}\gamma^{0})\psi_{\sigma}+\Delta^{*}(\psi^{T}_{\uparrow}Ci\gamma_{5}\psi_{\downarrow})+\Delta(\bar{\psi}_{\downarrow}i\gamma_{5}C\bar{\psi}^{T}_{\uparrow}).
\end{eqnarray}
To define finite-temperature Heisenberg operators, we take $\tau=i t$ and $x=(\tau,\mathbf{x})$. Similar to the BCS theory of superconductivity, here we introduce a Nambu-Gorkov spinor following Ref.~\cite{ZhuangPRD07a}:
\begin{equation}
\Psi=
\begin{pmatrix}\displaystyle
\psi_{\uparrow}\\
C\bar{\psi}^{T}_{\downarrow}
\end{pmatrix}
,\mbox{ }\bar{\Psi}=(\bar{\psi}_{\uparrow},\mbox{ }\psi^{T}_{\downarrow}C).
\end{equation}
We also introduce a Green's function matrix
\begin{eqnarray}
& &S(x,x^{\prime})=-\langle T_{\tau}[\Psi(x)\bar{\Psi}(x^{\prime})]\rangle .
\end{eqnarray}
Here $T_{\tau}$ denotes $\tau$-order.
From equations of motion one has
\begin{equation}
D(x)S(x,x^{\prime})=\mathbf{1}_{8\times8}\delta(x-x^{\prime}),
\end{equation}
where $\delta(x-x^{\prime})=\delta(\tau-\tau^{\prime})\delta(\mathbf{x}-\mathbf{x}^{\prime})$, $\mathbf{1}_{8\times8}$ is the eight-dimensional identity matrix, and $D(x)$ is a differential operator matrix. Its expression can be found in Appendix~\ref{app:a}. In momentum space, one has
\begin{equation}
D(K)S(K)=\mathbf{1_{8\times8}},
\end{equation}
Here $K=(i\omega_n,\mathbf{k})$ is a four-momentum at finite temperature with $\omega_n=(2n+1)\pi/ \beta$ denoting the fermion Matsubara frequency. Here we set $k_{B}\equiv 1$ and therefore $\beta=1/T$. We present our formalisms at general temperatures but only consider results at $T=0$.

The inverses of $D(K)$ corresponds to interacting fermion propagator matrix 
\begin{equation}
S(K)=
\left[ \begin{array} {ccc}
G_{\uparrow}(K,\mu_{\uparrow},-\mu_{\downarrow}) & F_{\uparrow\downarrow}(K,\mu_{\uparrow},-\mu_{\downarrow}) \\
F_{\downarrow\uparrow}(K,-\mu_{\downarrow},\mu_{\uparrow}) & G_{\downarrow}(K,-\mu_{\downarrow},\mu_{\uparrow})
\end{array} \right],
\end{equation}
where the matrix elements are
\begin{equation}
G_{\uparrow}(K,\mu_{\uparrow},-\mu_{\downarrow})=[G^{-1}_{0\uparrow}(K,\mu_{\uparrow})-\Sigma_{\uparrow}(K,-\mu_{\downarrow})]^{-1},
\end{equation}
\begin{equation}
F_{\uparrow\downarrow}(K,\mu_{\uparrow},-\mu_{\downarrow})=-G_{\uparrow}(K,\mu_{\uparrow},-\mu_{\downarrow})i\Delta\gamma_{5}G_{0\downarrow}(K,-\mu_{\downarrow}),
\end{equation}
Similar expressions for $G_{\downarrow}(K,\mu_{\downarrow},-\mu_{\uparrow})$ and $F_{\downarrow\uparrow}(K,\mu_{\downarrow},-\mu_{\uparrow})$ can be found. The fermion self-energy $\Sigma_{\sigma}$ is defined as
\begin{eqnarray}
\Sigma_{\uparrow}(K,-\mu_{\downarrow})=i\Delta\gamma_{5}G_{0\downarrow}(K,-\mu_{\downarrow})i\Delta\gamma_{5}
\end{eqnarray}
A similar expression for $\Sigma_{\downarrow}(K,-\mu_{\uparrow})$ can be found. The energy dispersions of relativistic fermions and antifermions are $\xi^{\pm}_{k\sigma}=\epsilon_{k}\pm\mu_{\sigma}$, where $\epsilon_{k}=\sqrt{\mathbf{k}^{2}+m^{2}}$. Here ``$-$'' is for fermions and ``$+$'' is for antifermions. The Fermi energy $E_{F}=\sqrt{k^{2}_{F}+m^{2}}$ and the Fermi momentum $k_{F}$ of an unpolarized non-interacting Fermi gas with equal total number density are chosen as units of energy and momentum. Here $n=k^{3}_{F}/(3\pi^{2})$ and $n$ denotes total number density.

We introduce two energy projectors
\begin{equation}
\Lambda_{\pm}(\mathbf{k})=\frac{1}{2}\left[1\pm\frac{\gamma^{0}(\vec{\gamma}\cdot\mathbf{k}+m)}{\epsilon_{k}}\right].
\end{equation}
By using these energy projectors, the non-interacting fermion propagator matrices can be written as 
\begin{equation}
G_{0\uparrow}(K,\mu_{\uparrow})=\left[\frac{\Lambda_{+}(\mathbf{k})}{i\omega_n-\xi^{-}_{k\uparrow}}+\frac{\Lambda_{-}(\mathbf{k})}{i\omega_n+\xi^{+}_{k\uparrow}}\right]\gamma^{0},
\end{equation} 
and a similar expression for $G_{0\downarrow}(K,\mu_{\downarrow})$. One can see that propagator of each species has contributions from both fermions and antifermions. The interacting fermion propagator matrices can be expressed as
\begin{eqnarray}
G_{\uparrow}(K,\mu_{\uparrow},-\mu_{\downarrow})&=&\Big[\frac{u^{(-)2}_{k}\Lambda_{+}(\mathbf{k})}{i\omega_n-E^{-}_{k\alpha}}+\frac{v^{(-)2}_{k}\Lambda_{+}(\mathbf{k})}{i\omega_n+E^{-}_{k\beta}} \nonumber \\
& &+\frac{u^{(+)2}_{k}\Lambda_{-}(\mathbf{k})}{i\omega_n+E^{+}_{k\alpha}}+\frac{v^{(+)2}_{k}\Lambda_{-}(\mathbf{k})}{i\omega_n-E^{+}_{k\beta}}\Big]\gamma^{0}.
\end{eqnarray}
A similar expression can be found for $G_{\downarrow}(K,\mu_{\downarrow},-\mu_{\uparrow})$. In the above expressions we introduced energy spectra for fermions, $E^{-}_{k\alpha,\beta}=E^{-}_{k}\pm\xi^{-}_{kb}$, and for antifermions, $E^{+}_{k\alpha,\beta}=E^{+}_{k}\pm\xi^{+}_{kb}$. Here we define $E^{-}_{k}=\sqrt{\xi^{-2}_{ka}+\Delta^{2}}$, $E^{+}_{k}=\sqrt{\xi^{+2}_{ka}+\Delta^{2}}$, $\xi^{\pm}_{ka}=\epsilon_{k}\pm\frac{\mu_{\uparrow}+\mu_{\downarrow}}{2}$, $\xi^{\pm}_{kb}=\pm\frac{\mu_{\uparrow}-\mu_{\downarrow}}{2}$, $u^{(\pm)2}_{k}=\frac{1}{2}(1+\frac{\xi^{\pm}_{ka}}{E^{\pm}_{k}})$ and $v^{(\pm)2}_{k}=\frac{1}{2}(1-\frac{\xi^{\pm}_{ka}}{E^{\pm}_{k}})$. Unlike the case of Fermi gases with only one species, here both energy spectra of fermions and antifermions may cross zero when there is a population asymmetry in the two-species. As a consequence, there may exist gapless excitations for both fermions and antifermions. We define $\mu=(\mu_{\uparrow}+\mu_{\downarrow})/2$ and $h=(\mu_{\uparrow}-\mu_{\downarrow})/2$. We choose the spin-up species as the majority so $h>0$. There are  zeros in energy spectra of fermions and antifermions. Zeros of the energy spectrum of fermions, if exist, are at $k^{2}_{1,2}=\mbox{max}(0,(\mu\pm\sqrt{h^{2}-\Delta^{2}})^{2}-m^{2})$. Zeros of the energy spectrum of antifermions, if exist, are at $k^{2}_{3,4}=\mbox{max}(0, (-\mu\pm\sqrt{h^{2}-\Delta^{2}})^{2}-m^{2})$. These zeros $k_{1,2,3,4}$ may not exist at the same time, but it can be shown that at least one of them must exist if there is a population asymmetry. As a consequence, a meaningful solution should have $|h|>\Delta$. It is also possible that the energy spectrum of fermions and that of antifermions may have different numbers of zeros. We also observe that the number of zeros in the energy spectrum of either fermions or antifermions may change. It is interesting that the topology of gapless regimes changes with the number of zeros in the spectra. However, this does not introduce any new phase or singularity.

Number density of each species is determined by $n_{\sigma}=\frac{1}{2}\sum_{K}\mbox{Tr}[\gamma^{0}G_{\sigma}(K)]$. After summing fermion Matsubara frequencies and making use of $\mbox{Tr}(\gamma^{0}\Lambda_{\pm}(\mathbf{k})\gamma^{0})=2$, one has
\begin{eqnarray}\label{eq:neq}
& &n_{\uparrow}=\sum_{\mathbf{k}}\big[u^{(-)2}_{k}f(E^{-}_{k\alpha})+v^{(-)2}_{k}f(-E^{-}_{k\beta})-(u^{(+)2}_{k}f(E^{+}_{k\alpha})+v^{(+)2}_{k}f(-E^{+}_{k\beta}))\big], \nonumber \\
& &n_{\downarrow}=\sum_{\mathbf{k}}\big[u^{(-)2}_{k}f(E^{-}_{k\beta})+v^{(-)2}_{k}f(-E^{-}_{k\alpha})-(u^{(+)2}_{k}f(E^{+}_{k\beta})+v^{(+)2}_{k}f(-E^{+}_{k\alpha}))\big],\nonumber \\
\end{eqnarray}
from which on can determine contributions from fermions and antifermions. Here $f(x)=[\exp(x/T)+1]$ is the Fermi distribution function. The polarization $p$ is defined as $p\equiv (n_{\uparrow}-n_{\downarrow})/(n_{\uparrow}+n_{\downarrow})$.

The $F$ functions can be expressed as
\begin{eqnarray}
F_{\uparrow\downarrow}(K,\mu_{\uparrow},-\mu_{\downarrow})&=&\frac{i\Delta}{2}\Big[\frac{\Lambda_{+}(\mathbf{k})}{E^{-}_{k}}\Big(\frac{1}{i\omega_n-E^{-}_{k\alpha}}-\frac{1}{i\omega_n+E^{-}_{k\beta}}\Big) \nonumber \\
& &+\frac{\Lambda_{-}(\mathbf{k})}{E^{+}_{k}}\Big(\frac{1}{i\omega_n-E^{+}_{k\beta}}-\frac{1}{i\omega_n+E^{+}_{k\alpha}}\Big)\Big]\gamma_{5}. 
\end{eqnarray}
A similar expression can be found for $F_{\downarrow\uparrow}(K,\mu_{\downarrow},-\mu_{\uparrow})$. From the definition of gap function, one can derive the gap equation.
\begin{eqnarray}
\Delta=-\frac{i}{2}g\sum_{K}\mbox{Tr}[\gamma_{5}F_{\uparrow\downarrow}(K,-\mu_{\uparrow},\mu_{\downarrow})]=-\frac{i}{2}g\sum_{K}\mbox{Tr}[\gamma_{5}F_{\downarrow\uparrow}(K,-\mu_{\downarrow},\mu_{\uparrow})].
\end{eqnarray}
Explicitly, 
\begin{equation}\label{eq:geq}
-\frac{1}{g}=\sum_{\mathbf{k}}\Big[\frac{1-f(E^{-}_{k\alpha})-f(E^{-}_{k\beta})}{2E^{-}_{k}}+\frac{1-f(E^{+}_{k\alpha})-f(E^{+}_{k\beta})}{2E^{+}_{k}}\Big].
\end{equation}

We present another equivalent derivation of the gap function. 
There are two possibilities for the asymmetric pair susceptibility in the presence of a number density asymmetry. They are defined as 
\begin{equation}
\chi_{\uparrow\downarrow}(Q)=\frac{1}{2}\sum_{K}\mbox{Tr}[G_{\uparrow}(K,\mu_{\uparrow},-\mu_{\downarrow})G_{0\downarrow}(Q-K,\mu_{\downarrow})]
\end{equation}
and $\chi_{\downarrow\uparrow}(Q)$ with a similar expression. 
When $Q=0$, it can be shown that 
\begin{equation}
\chi_{\uparrow\downarrow}(0)=\frac{1}{2\Delta}\sum_{K}\mbox{Tr}[i\gamma_{5}F_{\uparrow\downarrow}(K,\mu_{\uparrow},-\mu_{\downarrow})],
\end{equation}
and $\chi_{\downarrow\uparrow}(0)$ has a similar expression. Furthermore, one already saw that these two expressions at $Q=0$ coincide. The gap equation follows Thouless criterion \cite{Thoulesscriterion} which is given by $1+g\chi_{\uparrow\downarrow}(0)=0$ or $1+g\chi_{\downarrow\uparrow}(0)=0$. They are equivalent to Eq.~(\ref{eq:geq}).

Since our simple model is not renormalizable, a proper regularization is required. Following Ref.~\cite{ZhuangPRD07a}, the coupling constant is regularized by 
\begin{equation}
\frac{1}{g}=\frac{m}{4\pi a}-\frac{1}{2}\sum_{\mathbf{k}}\Big(\frac{1}{\epsilon_{k}-m}+\frac{1}{\epsilon_{k}+m}\Big),
\end{equation}
where $a$ is the $s$-wave scattering length. We introduce a momentum cutoff $\Lambda$ because there is an ultraviolet divergence.

The nonrelativistic limit is defined as the limit when $k_{F\sigma}\ll m$, $|\mu_{\sigma}-m|\ll m$, and $\Delta\ll m$. In this limit, all terms related to antifermion spectra or negative-energy solutions in the fermion propagator can be neglected. The energy spectrum of fermions becomes $\xi_{k\sigma}=\frac{\mathbf{k}^{2}}{2m}-(\mu_{\sigma}-m)$ and the fermion propagator is reduced to the usual non-relativistic results (for example, see \cite{Chien06}). Number densities also recover the nonrelativisitic results \cite{Chien06} if contributions from antifermions are dropped.

The ground state energy of the system is given by
\begin{eqnarray}
E_{S}
&=&\frac{1}{2}\sum_{K}\mbox{Tr}\Big\{\big[\gamma^{0}\epsilon_{k}\Lambda_{+}(\mathbf{k})-\gamma^{0}\epsilon_{k}\Lambda_{-}(\mathbf{k})+\frac{1}{2}\Sigma_{\uparrow}(K,-\mu_{\downarrow})\big]G_{\uparrow}(K,\mu_{\uparrow},-\mu_{\downarrow}) \nonumber \\
& &+\big[\gamma^{0}\epsilon_{k}\Lambda_{+}(\mathbf{k})-\gamma^{0}\epsilon_{k}\Lambda_{-}(\mathbf{k})+\frac{1}{2}\Sigma_{\downarrow}(K,-\mu_{\uparrow})\big]G_{\downarrow}(K,\mu_{\downarrow},-\mu_{\uparrow})\Big\} \nonumber \\
&=&\sum_{\mathbf{k},\lambda=\pm}\big[\xi^{\lambda}_{ka}-E^{\lambda}_{k}+E^{\lambda}_{k\alpha}f(E^{\lambda}_{k\alpha})+E^{\lambda}_{k\beta}f(E^{\lambda}_{k\beta})\big]+\mu n+h\delta n-\frac{\Delta^{2}}{g}.
\end{eqnarray}
The electrostatic energy is not included in our consideration.
 From the ground state energy one can derive the thermodynamical potential $\Omega_{S}\equiv E_{S}-TS-\mu_{\uparrow}n_{\uparrow}-\mu_{\downarrow}n_{\downarrow}$, which is 
\begin{eqnarray}\label{eq:OmegaS}
\Omega_{S}=-\frac{\Delta^{2}}{g}+\sum_{\mathbf{k},\lambda=\pm}\big[\xi^{\lambda}_{ka}-E^{\lambda}_{k}-T\ln(1+e^{-\frac{E^{\lambda}_{k\alpha}}{T}})-T\ln(1+e^{-\frac{E^{\lambda}_{k\beta}}{T}})\big].
\end{eqnarray}
It can also be shown that 
\begin{eqnarray}
n=-\Big(\frac{\partial\Omega_{S}}{\partial\mu}\Big)_{\Delta,h}, \qquad \delta n=-\Big(\frac{\partial\Omega_{S}}{\partial h}\Big)_{\Delta,\mu}
\end{eqnarray}
give the same number equations and 
\begin{equation}
\Big(\frac{\partial\Omega_{S}}{\partial\Delta}\Big)_{\mu,h}=0 
\end{equation}
 gives the same gap equation. Our expression for $\Omega_{S}$ is consistent with the result in Ref.~\cite{Chatterjee08} although in that work a variational ansatz with a coherent state is implemented to obtain $\Omega_{S}$.

\subsection{Bogoliubov Canonical Transformation}
The results obtained so far can also be derived from a canonical transformation. In previous subsections, our derivation follows conventional BCS ansatz, in which the ground state is assumed to be the BCS-Leggett ground state. Since the energy spectrum of quasi-fermions may become gapless, quasi-fermions survive in those regimes at zero temperature. In the deep BEC side when attraction is strong, the energy spectrum of quasi-antifermions can also be gapless. Thus in those regime quasi-antifermions can be found in the ground state. This is a new phenomenon which cannot be observed in nonrelativistic systems. It is presumed that in BCS theory that fluctuations of the quantum operator $g\psi^{T}_{\uparrow}Ci\gamma_{5}\psi_{\downarrow}$ around its mean field value are negligible. As a consequence, the Hamiltonian derived from Eq.~(\ref{eq:LMF}) can be linearized. It becomes
\begin{eqnarray}
H(x)&=&\sum_{\sigma=\uparrow,\downarrow}\bar{\psi}_{\sigma}(-i\vec{\gamma}\cdot\nabla+m-\mu_{\sigma}\gamma^{0})\psi_{\sigma} \nonumber \\
& &-\Delta^{*}(\psi^{T}_{\uparrow}Ci\gamma_{5}\psi_{\downarrow})-\Delta(\bar{\psi}_{\downarrow}i\gamma_{5}C\bar{\psi}^{T}_{\uparrow})-\frac{\Delta^{2}}{g}.
\end{eqnarray}
The order parameter $\Delta$ can be taken as real with no loss of generality. Introducing annihilation and creation operators for fermions and antifermions, $a_{\sigma}(\mathbf{k})$, $a^{\dagger}_{\sigma}(\mathbf{k})$, $b_{\sigma}(\mathbf{k})$ and $b^{\dagger}_{\sigma}(\mathbf{k})$, the fermion field operator can be expanded in the momentum space as 
\begin{eqnarray}
\psi_{\sigma}(x)&=&\sum_{\mathbf{k}}\frac{1}{\sqrt{2\epsilon_{k}}}\Bigg\{
\left[ \begin{array}{ccc}
\sqrt{\epsilon_{k}+k}\eta^{L}_{\mathbf{k}} \\
\sqrt{\epsilon_{k}-k}\eta^{L}_{\mathbf{k}}
\end{array}\right]a_{\sigma}(\mathbf{k})e^{-ik\cdot x} \nonumber \\
& &+\left[ \begin{array}{ccc}
\sqrt{\epsilon_{k}-k}\eta^{R}_{\mathbf{k}} \\
-\sqrt{\epsilon_{k}+k}\eta^{R}_{\mathbf{k}}
\end{array}\right]b^{\dagger}_{\sigma}(\mathbf{k})e^{ik\cdot x}\Bigg\},
\end{eqnarray}
where $\eta^{L}_{\mathbf{k}}$ and $\eta^{R}_{\mathbf{k}}$ correspond to two eigenvectors of $\vec{\sigma}\cdot\hat{\mathbf{k}}$ with different handedness and $k_{\mu}=(\epsilon_{k},\mathbf{k})$.
\begin{equation}
\eta^{L}_{\mathbf{k}}=
\left[ \begin{array}{ccc}
-\mbox{sin}\frac{\theta_{\mathbf{k}}}{2}e^{-i\phi_{\mathbf{k}}} \\
\mbox{cos}\frac{\theta_{\mathbf{k}}}{2}
\end{array}\right] ,\qquad  \eta^{R}_{\mathbf{k}}=
\left[ \begin{array}{ccc}
\mbox{cos}\frac{\theta_{\mathbf{k}}}{2} \\
\mbox{sin}\frac{\theta_{\mathbf{k}}}{2}e^{i\phi_{\mathbf{k}}}
\end{array}\right].
\end{equation}
Here $\theta$ and $\phi$ are polar and azimuthal angels of the vector $\mathbf{k}$. One can write the Hamiltonian in terms of $a$'s and $b$'s. To make the result more compact, we redefine $a_{\sigma}(\mathbf{k})=e^{i\frac{\phi_{\mathbf{k}}+\pi}{2}}\bar{a}_{\sigma}(\mathbf{k})$ and $b_{\sigma}(\mathbf{k})=e^{i\frac{\phi_{\mathbf{k}}}{2}}\bar{b}_{\sigma}(\mathbf{k})$, the Hamiltonian finally becomes 
\begin{eqnarray} 
& &H=-\frac{\Delta^{2}}{g}+\sum_{\mathbf{k}}\Big\{\sum_{\sigma=\uparrow,\downarrow}\big[\xi^{-}_{k\sigma}\bar{a}^{\dagger}_{\sigma}(\mathbf{k})\bar{a}_{\sigma}(\mathbf{k})+\xi^{+}_{k\sigma}\bar{b}^{\dagger}_{\sigma}(\mathbf{k})\bar{b}_{\sigma}(\mathbf{k})\big] \nonumber \\
& &-\Delta\big[\bar{a}_{\downarrow}(-\mathbf{k})\bar{a}_{\uparrow}(\mathbf{k})+\bar{b}^{\dagger}_{\uparrow}(\mathbf{k})\bar{b}^{\dagger}_{\downarrow}(-\mathbf{k})+\bar{a}^{\dagger}_{\uparrow}(\mathbf{k})\bar{a}^{\dagger}_{\downarrow}(\mathbf{k})+\bar{b}_{\downarrow}(-\mathbf{k})\bar{b}_{\uparrow}(\mathbf{k})\big]\Big\}.
\end{eqnarray}
Now we introduce the BCS-Leggett ground state for relativistic systems 
\begin{equation}
|\mbox{BCS}\rangle=\prod_{\mathbf{k}}[u^{-}_{k}+v^{-}_{k}\bar{a}^{\dagger}_{\uparrow}(\mathbf{k})\bar{a}^{\dagger}_{\downarrow}(-\mathbf{k})][u^{+}_{k}+v^{+}_{k}\bar{b}^{\dagger}_{\uparrow}(\mathbf{k})\bar{b}^{\dagger}_{\downarrow}(-\mathbf{k})]|0\rangle
\end{equation}
and a Bogoliubov canonical transformation 
\begin{eqnarray}
& &\begin{pmatrix}\displaystyle
\alpha(\mathbf{k}) \\
\beta^{\dagger}(\mathbf{k}) 
\end{pmatrix}=
\begin{pmatrix}\displaystyle
u^{-}_{k} & -v^{-}_{k} \\
v^{-}_{k} & u^{-}_{k}
\end{pmatrix}
\begin{pmatrix}\displaystyle
\bar{a}_{\uparrow}(\mathbf{k}) \\
\bar{a}^{\dagger}_{\downarrow}(-\mathbf{k})
\end{pmatrix}, \nonumber \\
& &\begin{pmatrix}\displaystyle
\bar{\alpha}(\mathbf{k}) \\
\bar{\beta}^{\dagger}(\mathbf{k}) 
\end{pmatrix}=
\begin{pmatrix}\displaystyle
u^{+}_{k} & -v^{+}_{k} \\
v^{+}_{k} & u^{+}_{k}
\end{pmatrix}
\begin{pmatrix}\displaystyle
\bar{b}_{\uparrow}(\mathbf{k}) \\
\bar{b}^{\dagger}_{\downarrow}(-\mathbf{k})
\end{pmatrix}.
\end{eqnarray}
It can be shown that $\alpha(\mathbf{k})|\mbox{BCS}\rangle=0$ and $\beta(-\mathbf{k})|\mbox{BCS}\rangle=0$. Therefore $\alpha(\mathbf{k})$ and $\beta(-\mathbf{k})$ are annihilation operators of a quasi-fermion and a quasi-antifermion, respectively. 
The Hamiltonian can be diagonalized following this transformation. It becomes
\begin{eqnarray}
& &H=-\frac{\Delta^{2}}{g}+\sum_{\mathbf{k}}\Big(\xi^{-}_{ka}-E^{-}_{k}+\xi^{+}_{ka}-E^{+}_{k}+E^{-}_{k\alpha}\alpha^{\dagger}(\mathbf{k})\alpha(\mathbf{k}) \nonumber \\
& &\qquad+E^{-}_{k\beta}\beta^{\dagger}(\mathbf{k})\beta(\mathbf{k})+E^{+}_{k\alpha}\bar{\alpha}^{\dagger}(\mathbf{k})\bar{\alpha}(\mathbf{k})+E^{+}_{k\beta}\bar{\beta}^{\dagger}(\mathbf{k})\bar{\beta}(\mathbf{k})\Big).
\end{eqnarray}
The order parameter in terms of this new set of operators has the following form 
\begin{equation}\label{eq:Bogogap}
\Delta=-g\sum_{\mathbf{k}}\langle\bar{a}_{\downarrow}(-\mathbf{k})\bar{a}_{\uparrow}(\mathbf{k})+\bar{b}^{\dagger}_{\uparrow}(\mathbf{k})\bar{b}^{\dagger}_{\downarrow}(-\mathbf{k})\rangle.
\end{equation}
With number densities for quasi-fermions $\langle\alpha^{\dagger}(\mathbf{k})\alpha(\mathbf{k})\rangle=f(E^{-}_{k\alpha})$ and $\langle\beta^{\dagger}(\mathbf{k})\beta(\mathbf{k})\rangle=f(E^{-}_{k\beta})$ and those for antiquasifermions, it can be shown that Eq.~(\ref{eq:Bogogap}) is equivalent to Eq.~(\ref{eq:geq}). Similarly, number densities $n_{\sigma}=\langle\psi^{\dagger}_{\sigma}\psi_{\sigma}\rangle$ give the same number equations as Eq.~(\ref{eq:neq}). 
From the diagonalized Hamiltonian, one can calculate the thermodynamical potential $\Omega_{S}=-\beta^{-1}\ln\mbox{Tr}\left(e^{-\beta H}\right),$ 
which has exactly the same expression as Eq.~(\ref{eq:OmegaS}). Therefore these two formalisms are equivalent. We would like to point out that in the presence of inhomogeneity such as an external potential, the Bogoliubov canonical transformation can be generalized via the Bogoliubov - de Gennes equation \cite{DG66} to study these inhomogeneous phases. This will be particularly useful in the study of modulating phases such as the LOFF state.

\section{Stability Condition and Phase Separation}\label{sec:stab}
After solving the number and gap equations consistently, one obtains a homogeneous phase. However, it may not be stable. To check its stability, we analyze the number susceptibility matrix $\partial n_{\sigma}/\partial\mu_{\sigma^{\prime}}$, which is related to the compressibility. The number susceptibility matrix reveals responses of the system to small perturbations in densities. Only when all eigenvalues of the number susceptibility matrix are non-negative, the phase is stable against phase separation in real space \cite{Stability,Pao}. This is the mechanical stability criterion. For a charged Fermi gas, there is another stability criterion which requires that the sqaured Meissner mass of the gauge bosons be positive. This additional stability criterion is discussed in \cite{HuangPRD04} for gapless color superconducting states. We only present results from the mechanical stability criterion.

The gap function $\Delta$ is an implicit function of $\mu_{\sigma^{\prime}}$ (or $\mu$ and $h$) if the gap equation is satisfied. For simplicity in notation, we use the ordinary partial derivative $\frac{\partial}{\partial x}$ when $\Delta$, $\mu$ and $h$ are all treated as independent variables, and use $\frac{D}{Dx}\equiv\frac{\partial}{\partial x}+\frac{\partial \Delta}{\partial x}\frac{\partial}{\partial \Delta}$ $(x=\mu,h)$ as the total derivative with respect to $\mu$ and $h$ when $\Delta$ can be inferred from the gap equation. The number susceptibility matrix can be written as 
\begin{equation}
\begin{pmatrix}\displaystyle
\frac{Dn_{\uparrow}}{D\mu_{\uparrow}} & \displaystyle \frac{Dn_{\uparrow}}{D\mu_{\downarrow}} \\
\displaystyle \frac{Dn_{\downarrow}}{D\mu_{\uparrow}} & \displaystyle \frac{Dn_{\downarrow}}{D\mu_{\downarrow}}
\end{pmatrix} 
=\frac{1}{2} A \begin{pmatrix}
\displaystyle \frac{Dn}{D\mu} & \displaystyle \frac{Dn}{Dh} \\
\displaystyle \frac{D\delta n}{D\mu} & \displaystyle \frac{D\delta n}{Dh}
\end{pmatrix}  A  = \frac{1}{2}AMA,
\end{equation}
where $A=A^{-1}=\frac{\sqrt{2}}{2} \begin{pmatrix} 1 & 1 \\ 1 & -1 \end{pmatrix}$. Therefore, eigenvalues of the matrix are given by $\lambda_\pm=\left[\mbox{Tr}(M)\pm\sqrt{\mbox{Tr}(M)^2 - 4\mbox{det}(M)}\right]/2$, where $\mbox{Tr}(M)$ and $\mbox{det}(M)$ denote the trace and determinant of $M$, respectively. Matrix elements of $M$ are list in Appendix~\ref{app:b}.

If a homogeneous solution is not stable, it is possible that phase separation in real space or a LOFF state may take place. It is known that the LOFF state is not likely to exist in the strongly attractive regime for non-relativistic fermions \cite{SR06}. Here we only consider phase separation for neutral relativistic fermions. In a phase separated state, there is an interface separating a superfluid and a normal phase. In the following sections we will show that homogeneous superfluid phase is unstable in two regimes, one in the weakly attractive regime and one deep in the RBEC regime. Chiral condensates with $\langle\bar{\psi}\psi\rangle\neq 0$ which are not considered in our model may also occupy those unstable regimes.

Following \cite{Caldas03_04,ChienPRL}, we consider a phase separated state with volume fractions of normal and superfluid phases being $x$ and $(1-x)$, respectively. To achieve a stable phase separation, temperature, chemical potentials of the two species, and pressure should be continuous across the interface to balance heat convection, diffusion, and mechanical movement \cite{ChienPRL}. Explicitly, we require $\Omega_{N}=\Omega_{S}$ and $\mu_{N\sigma}=\mu_{\sigma}$. Here the thermodynamical potential of a normal phase is given by
\begin{equation}
\Omega_{N}=-T\sum_{\mathbf{k,\lambda=\pm}}\big[\ln(1+e^{-\frac{\xi^{\lambda}_{k\uparrow}}{T}})+\ln(1+e^{-\frac{\xi^{\lambda}_{k\downarrow}}{T}})\big].
\end{equation}
When electrostatic energy is taken into account, a phase separated state may have higher energy due to charge accumulation. This may disfavor the phase separated state and other inhomogeneous phases such as the LOFF phase. In that case, other homogeneous phases (such as chiral condensates) should be considered.

\section{Results and Discussions}\label{sec:result}
Following \cite{ZhuangPRD07a}, a BCS-BEC-RBEC crossover can be characterized by three dimensionless constants, the dimensionless coupling constant $\eta=1/(k_{F}a)$, $\zeta=k_{F}/m$, and the cutoff $z=\Lambda/(\zeta m)$. The quantity $\zeta$ describes how relativistic the system is. If $\zeta\ll1$, one should recover nonrelativistic results when the coupling constant is small. In this paper, we set $\Lambda/m=10$ and focus on two cases: $\zeta=0.1$ (weakly relativistic) and $\zeta=0.6$ (relativistic). When $\eta$ is extremely large, it is possible that $p>1$ \cite{Heprivate} as one can see from Eq.~(\ref{eq:neq}). Throughout this paper we consider $p\le 1$.

As mentioned in Sec.~\ref{sec:intro}, relativistic effects become significant when $\eta_{c}\sim\sqrt{2}\zeta^{-1}$ for a single-species system. Close to this critical value of coupling constant, $\eta_{c}$, a RBEC-NBEC crossover is expected. We observe that $\eta_{c}$ depends on $p$ in the presence of population imbalance. At zero temperature, we derive an analytical expression for the critical coupling constant. At $\eta_{c}$ , zeros of energy spectra of fermions and antifermions can be found at $k_{1}=k_{3}=0$ and $k_{2}\simeq k_{4}\simeq \sqrt{h^{2}-m^{2}}$ by using the approximations $\mu\simeq0$ and $\Delta\ll m$. It can also be shown that $p\simeq (k_{2}/k_{F})^{3}$. With these expressions, the critical coupling constant is found to be 
\begin{eqnarray}\label{eq:etac}
\eta_{c}
&=&\frac{2}{\pi}\Big[\zeta^{-1}\ln\Big(\frac{\Lambda}{m}+\sqrt{\big(\frac{\Lambda}{m}\big)^{2}+1}\Big)+\frac{p^{\frac{1}{3}}}{2}\sqrt{\zeta^{2}p^{\frac{2}{3}}+1}\nonumber \\
&-&\frac{\zeta^{3}}{2}\ln\Big(\zeta p^{\frac{1}{3}}+\sqrt{\zeta^{2}p^{\frac{2}{3}}+1}\Big)\Big].
\end{eqnarray}
From this expression, one sees that when $\zeta\ll1$ in the nonrelativistic regime the first term dominates and therefore $\eta_{c}$ is insensitive to $p$. For larger $\zeta$, the second term is no longer negligible and $\eta_{c}$ will have observable dependence on $p$.

\begin{figure}
\includegraphics[width=4.5in,clip]
{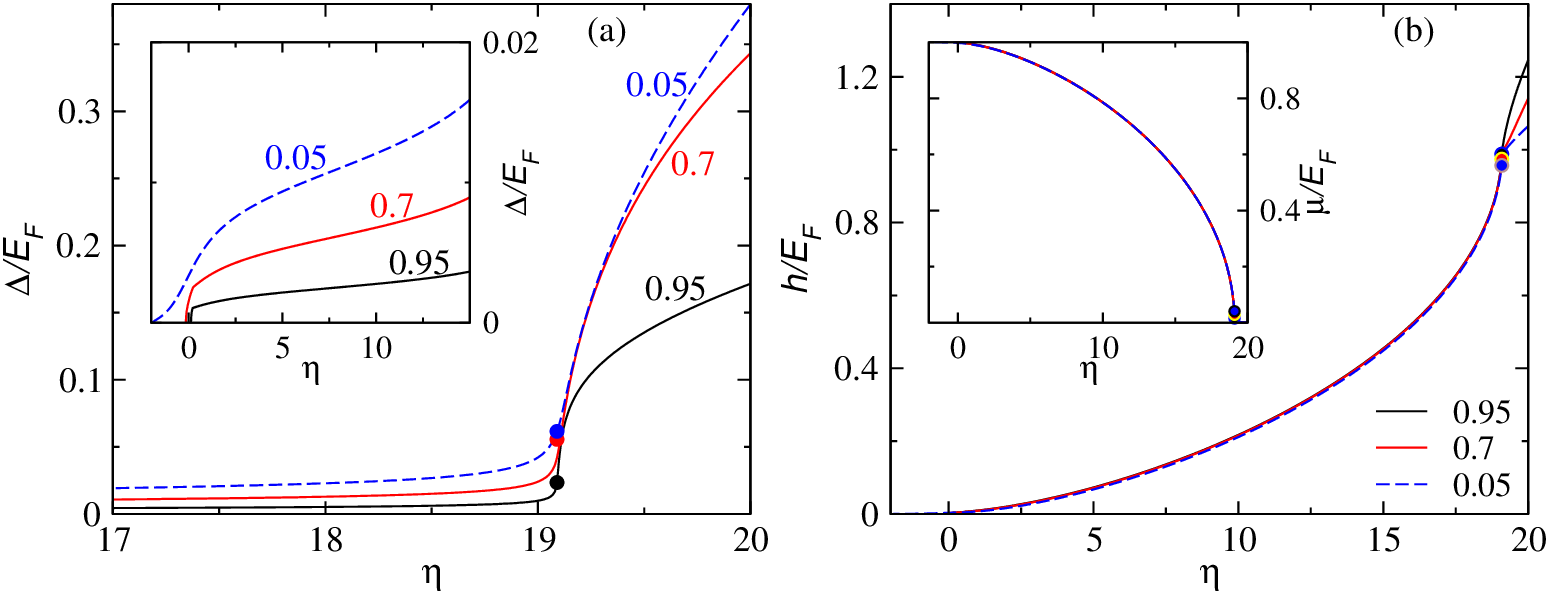}
\caption{(Color online) Signatures of a NBEC-RBEC crossover for $\zeta=0.1$. (a) $\Delta$ as a function of $\eta$ for selected values of polarization $p$ (labeled next to each curve). Inset: $\Delta$ in the weakly attractive regime. (b) $h$ and $\mu$ (inset) as a function of $\eta$ for selected values of polarization $p$ (labeled next to each curve). Solid dots indicate positions where relativistic effects become significant and crossovers take place.} 
\label{fig:Delta01}
\end{figure}

\begin{figure}
\includegraphics[width=4.5in,clip]
{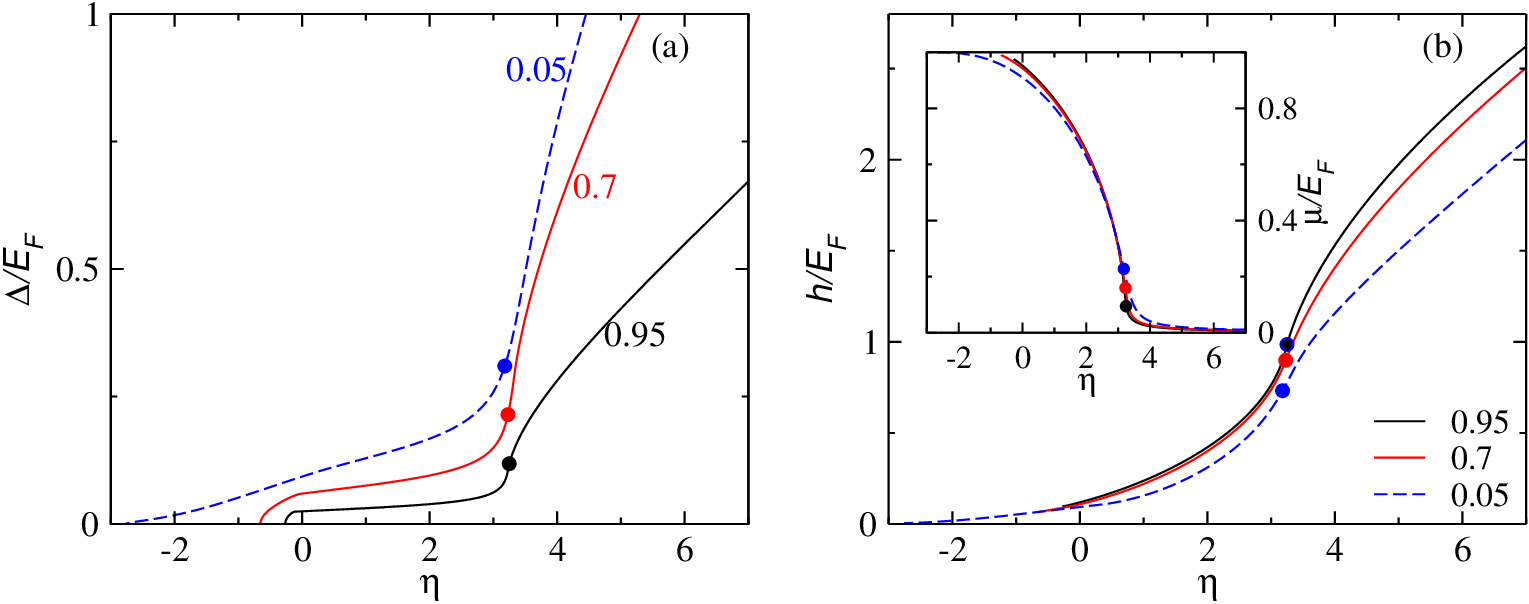}
\caption{(Color online) Signatures of a NBEC-RBEC crossover for $\zeta=0.6$. (a) $\Delta$ as a function of $\eta$ for selected values of polarization $p$ (labeled next to each curve). (b) $h$ and $\mu$ (inset) as a function of $\eta$ for selected values of $p$ (labeled next to each curve). Solid dots indicate positions where relativistic effects become significant and crossovers take place.} 
\label{fig:Delta06}
\end{figure}

\begin{figure}
\includegraphics[width=4.5in,clip]
{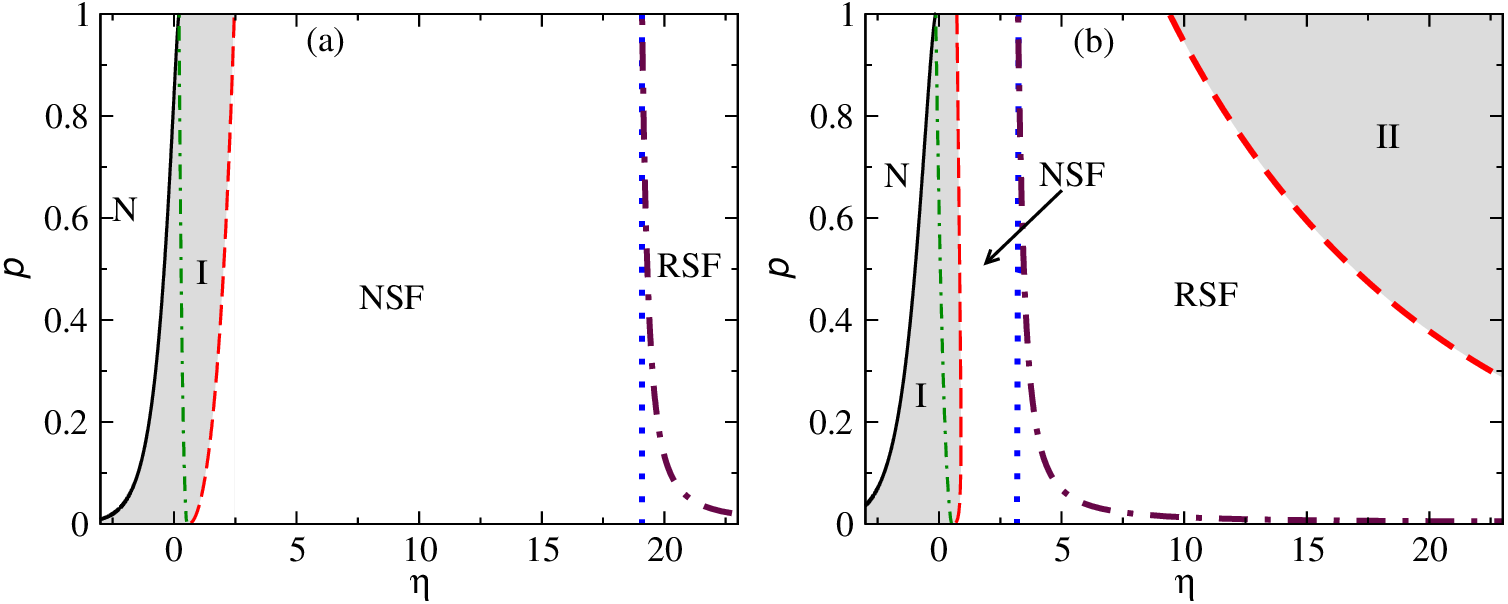}
\caption{(Color online) Phase diagrams at zero temperature for (a) $\zeta=0.1$ and (b) $\zeta=0.6$. Phase transitions occur only when one crosses solid lines. NSF and RSF denote nonrelativistic and relativistic superfluid phases, respectively. Shaded regimes correspond to unstable superfluids, which are nonrelativistic in regime I and are relativistic in regime II. Black sold lines indicate where $\Delta=0$, which are boundaries between normal phases (N) and USFs. Blue dashed lines indicate $\eta_{c}$, which is given by Eq.~(\ref{eq:etac}) and corresponds to the onset of RSF. Red dashed lines indicate boundaries between stable superfluids and unstable superfluids. Green dot-dash lines and maroon dot-dash lines indicate where the topology of gapless regimes in energy spectra of fermions and antifermions changes, respectively. To the right of the maroon dot-dash lines, gapless excitations of antifermions can be observed.} 
\label{fig:phase1}
\end{figure}

We demonstrate typical behavior of BCS-BEC-RBEC crossover in Fig.~\ref{fig:Delta01} (for $\zeta=0.1$) and Fig.~\ref{fig:Delta06} (for $\zeta=0.6$) by showing $\Delta$, $\mu$, and $h$ as a function of $\eta$ for selected values of $p$. Solid dots on those curves indicate where relativistic effects become significant. Their locations are obtained from Eq.~(\ref{eq:etac}). In both cases, $\Delta$, $\mu$ and $h$ change dramatically near $\eta_{c}$. $\Delta$ and $h$ become of order of $E_{F}$ while $\mu$ approaches zero (and fermions and antifermions become nearly degenerate). One can see that locations of those solid dots shift to smaller coupling constants when the system is more relativistic.  In both figures $\Delta$ decreases as $p$ increases, this is similar to the nonrelativistic case since the presence of number density asymmetry frustrate pairing. We also found that $\mu$ has weak dependence on $p$ while $h$ has more observable dependence on $p$ when $\eta>\eta_{c}$. The reason for the former is that the system has a large two-body binding energy when the coupling constant is large. Since the chemical potential is dominated by this binding energy which is independent of $p$,  the energy needed to break a bound pair ($\sim 2|\mu-m|$) has a very weak dependence on $p$. The reason for the latter is that when the coupling constant is large, excess quasi-fermions form a sphere in momentum space similar to a Fermi sphere of non-interacting fermions. $h$ is related to the radius of this sphere and therefore it is sensitive to $\delta n$, which is controlled by $p$. These features are more prominent in Fig.~\ref{fig:Delta01}.

\begin{figure}
\includegraphics[width=4.5in,clip]
{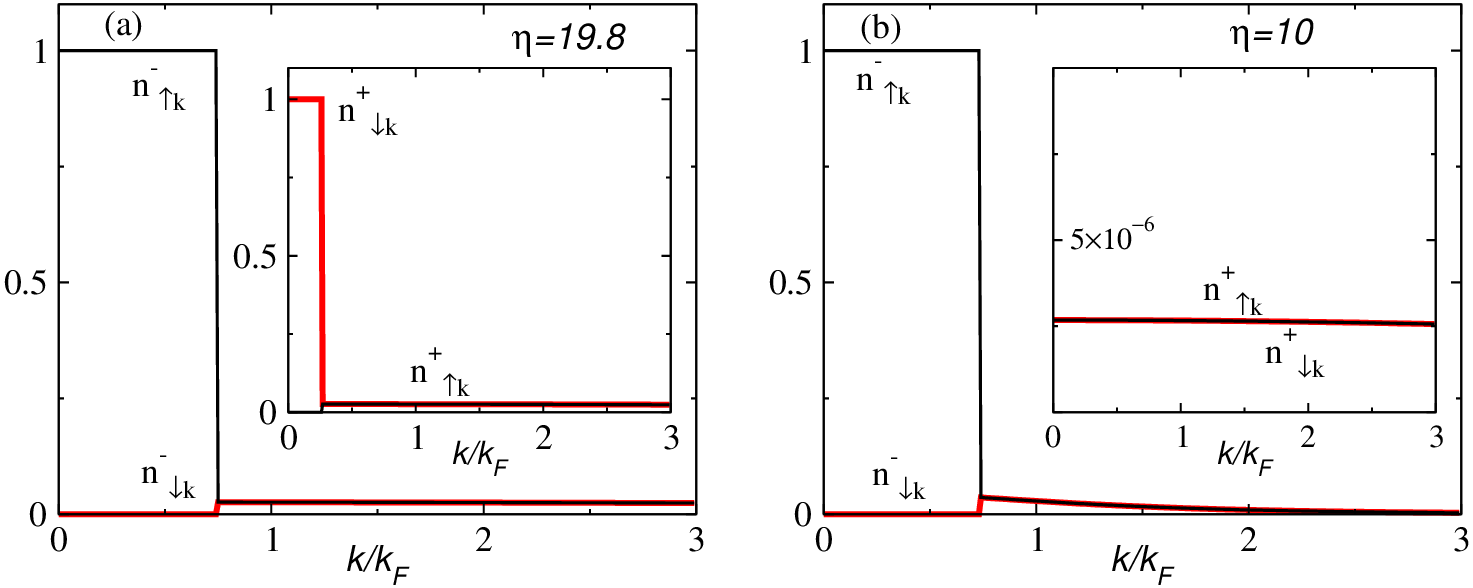}
\caption{(Color online) Number density distributions (in unit of $k_{F}^{3}$) for fermions as a function of momentum for $p=0.2$ and $\zeta=0.1$ with (a) $\eta=19.8$ (in the RBEC regime) and (b) $\eta=10$. Insets show corresponding number density distributions for antifermions. We use black and red to represent the two species.} 
\label{fig:density01}
\end{figure}

\begin{figure}
\includegraphics[width=4.5in,clip]
{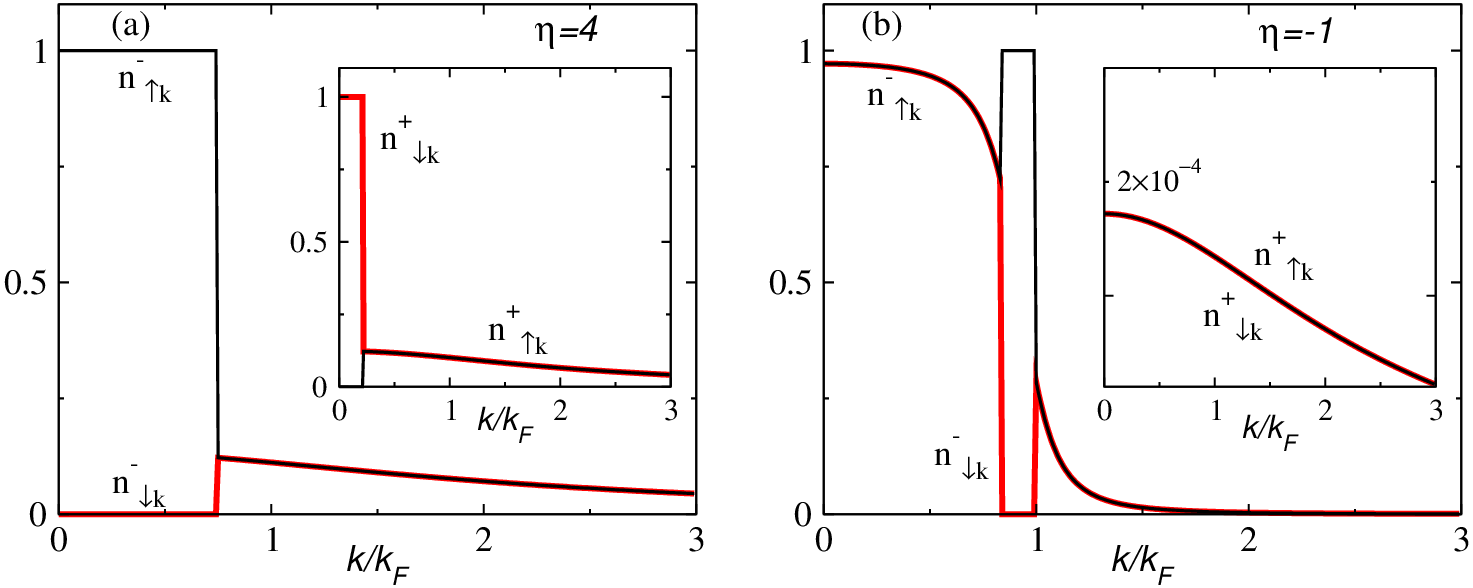}
\caption{(Color online) Number density distributions (in unit of $k_{F}^{3}$) for fermions as a function of momentum for $p=0.2$ and $\zeta=0.6$ with (a) $\eta=4$ (in the RBEC regime) and (b) $\eta=-1$. Insets show corresponding number density distributions for antifermions. We use black and red to represent the two species.} 
\label{fig:density06}
\end{figure}

Fig.~\ref{fig:phase1} shows a predictive phase diagrams in $p$-$\eta$ plane for (a) $\zeta=0.1$ and (b) $\zeta=0.6$. When $\zeta$ is small (weakly relativistic cases such as (a)), the phase diagram is similar to that of nonrelativistic cases, especially when $\eta$ is small. We show locations of $\eta_{c}$ on our phase diagrams. We found that crossing $\eta_{c}$ does not introduce any phase transition. Therefore a nonrelativistic superfluid (NSF) and a relativistic superfluid (RSF) are connected smoothly without singularities and we use dashed lines to indicate that only smooth crossovers occur. The dot-dash lines show where the topology of regimes of gapless quasi-fermions (green line) and quasi-antifermions (maroon line) changes. However, no new phase or singularities occur as the topology changes. We also show unstable regimes of homogeneous superfluids and our results agree with those presented in Ref.~\cite{WangPRD07}. We also test the stability of homogeneous superfluids. As shown in Fig.~\ref{fig:phase1}(b), there are two separated unstable regimes where number susceptibility matrix is not positive-semi-definite. In addition to being unstable when the coupling constant is small, a relativistic superfluid is also unstable when the coupling constant is very large.

In Fig.~\ref{fig:phase1}(a), the other unstable region of RSF appears when $\eta\gtrsim59.7$ and is not shown. We name the unstable regime with small values of $\eta$ as regime I and that with larger values of $\eta$ as regime II. Inside the regime I, there are gapless superfluid states with two zeros in the energy spectrum of quasifermions, which are similar to the gapless CFL or 2SC states \cite{instabCFL2SC1,HuangPRD04}.
The unstable regime II shifts deep into the RBEC side (with a larger coupling constant) as $\zeta$ decreases (or the system becomes more nonrelativistic). Therefore observations of this unstable regime II is unlikely in nonrelativistic systems.  
As $\zeta$ increases, the regime occupied by a nonrelativistic superfluid phase (NSF) shrinks. We found that it will eventually disappear completely at large enough $\zeta$ and only RSF phases survive. In charged Fermi gases, it is likely that inside regime I, the Meissner mass of gauge bosons may become imaginary, which indicates a new unstable regime \cite{ShovkovyPLB06}.

\begin{figure}
\includegraphics[width=4.5in,clip]
{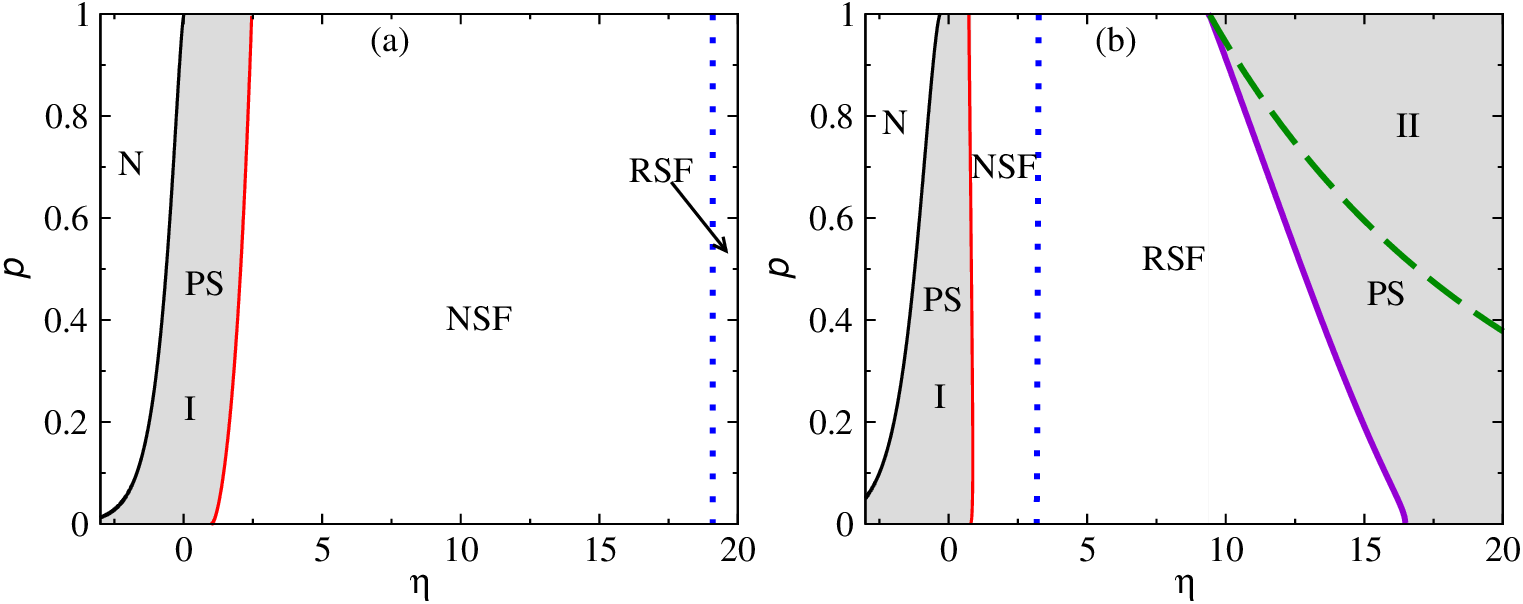}
\caption{(Color online) Phase diagrams including phase separation for (a) $\zeta=0.1$ and (b) $\zeta=0.6$. Shaded regimes labeled by PS correspond to phase separation. For phase separation covering the unstable regime I, black solid lines indicate boundaries between normal (N) phase and phase separation and red sold lines indicate boundaries between stable nonrelativistic superfluid (NSF) and phase separation. For phase separation covering unstable regime II, violet solid lines indicate boundaries between stable relativistic superfluid (RSF) and phase separation. Brown dot-dash lines indicate where $\Delta=0$ in unstable nonrelativistic superfluids. Blue and green dot-dash lines indicate where homogeneous superfluids become unstable. Orange dashed lines indicate $\eta_{c}$, which correspond to the onset of RSF.} 
\label{fig:phase2}
\end{figure}

We present number density distributions as a function of momentum for fermions and antifermions with $p=0.2$ in Fig.~\ref{fig:density01} for $\zeta=0.1$ and in Fig.~\ref{fig:density06} for $\zeta=0.6$. When $\eta>\eta_{c}$, antifermions develop a gapless regime in number density distributions. One can see that both fermions and antifermions are polarized. However, we observe an interesting feature: antifermions with spin-down has a larger population than antifermions with spin-up do, which is exactly opposite to the case of fermions. Due to this behavior of antifermions, the polarization of fermions needs to exceed the total polarization of the system. This is another distinct feature of relativistic fermionic systems. When the coupling constant is small, there is virtually no gapless excitation of antifermions. Comparisons between Fig.~\ref{fig:density01}(b) and Fig.~\ref{fig:density06}(b) show that the gapless regimes of quasi-fermions have different topologies since the gapless regime in Fig.~\ref{fig:density01}(b) has only one zero and the one in Fig.~\ref{fig:density06}(b) has two zeros.

In Fig.~\ref{fig:phase2} we present predictive phase diagrams including phase separation (PS) for (a) $\zeta=0.1$ and (b) $\zeta=0.6$. In previous sections, we found two regimes where homogeneous superfluid phases become unstable. We call them region I with small values of $\eta$ and region II with large values of $\eta$. In regime I, LOFF states or phase separation may appear and compete with each other. In region II, phases with chiral condensates ($\langle\bar{\psi}\psi\rangle\neq 0$) or phase separation may appear. 
In Fig.~\ref{fig:phase2}(a) only one of the two phase separation regimes is shown because the other one is too deep in the strongly attractive regime. For phase separation in regime I, phase separation covers roughly the regime where homogeneous superfluids become unstable. In contrast, in regime II phase separation covers a significantly larger regime than the regime where homogeneous superfluids become unstable. We also found that in regime II, the volume fraction of the normal phase of a phase separation only increases slowly with the couping constant $\eta$. For more complicated models with electrostatic energy, the boundaries of regimes where phase separation exists may shift.

\section{Conclusions}
In this paper we address relativistic BCS-BEC-RBEC crossover of fermions with tunable attractive interactions at zero temperature. We implement a BCS-Leggett ansatz and use two equivalent formalisms to investigate possible phases. We show that gapless excitations of quasi-fermions and quasi-antifermions are both possible. The stability of possible homogeneous polarized superfluid phases is analyzed. When homogeneous superfluid phases become unstable, we consider phase separation in real space and leave more exotic phases for future study. Number density distributions and predictive phase diagrams are presented to help understand effects of number density asymmetry on different phases in a relativistic BCS-BEC-RBEC crossover. Our results are consistent with the well-known results in the nonrelativistic limit.

It is of interest to extend our study to finite temperatures. It is conjectured that color superconductivity vanishes at high temperatures and the system becomes a quark gluon plasma. From study of BCS-BEC crossover, it is known that condensed Cooper pairs can either be broken into quasi-particles or be excited and become non-condensed pairs \cite{NSR}. To include both possibilities, one needs to consider pairing fluctuations and this is beyond mean field theories. In Ref.~\cite{ZhuangPRD07b} finite-temperature effects are considered for a one-species Fermi gas. It is possible to extend this approach to include finite-temperature effects for a two-species relativistic Fermi gas in future work.

This work is supported by Grant NSF PHY-0555325. We thank K. Levin for helpful discussions and P. Zhuang and L. He for useful comments.

\appendix
\section{Differential operator matrices} \label{app:a}
The differential operator matrix is 
\begin{equation}
D(x)=
\left[ \begin{array}{ccc}
D_{0\uparrow}(x,\mu_{\uparrow}) & i\Delta \gamma_{5} \\
i\Delta \gamma_{5} & D_{0\downarrow}(x,-\mu_{\downarrow}) 
\end{array} \right] .
\end{equation} 
Here $D_{0\sigma}(x,\mu_{\sigma})=-\gamma^{0}\partial_{\tau}+i\vec{\gamma}\cdot\nabla-(m-\mu_{\sigma}\gamma^{0})$ is the noninteracting differential operator matrix. In momentum space,  
\begin{equation}
D(K)=
\left[ \begin{array}{ccc}
G^{-1}_{0\uparrow}(K,\mu_{\uparrow}) & i\Delta \gamma_{5} \\
i\Delta \gamma_{5} & G^{-1}_{0\downarrow}(K,-\mu_{\downarrow})
\end{array} \right],
\end{equation} 
where $G^{-1}_{0\sigma}(K,\mu_{\sigma})=(i\omega_n+\mu_{\sigma})\gamma^{0}-\vec{\gamma}\cdot\mathbf{k}-m$ is the inverse of a non-interacting propagator.

\section{Number susceptibility matrix} \label{app:b}
The elements of the transformed number susceptibility matrix are given by the second-order derivatives of the thermodynamic potential $\Omega_{S}$
\begin{subequations}
\begin{eqnarray}
& &\frac{D n}{D\mu}  = 
- \frac{\partial^2\Omega_{S}}{\partial\mu^2} +
\left(\frac{\partial^2\Omega_{S}}{\partial\mu\partial\Delta}\right)^2 /\frac{\partial^2
  \Omega_{S}}{\partial\Delta^2} \,,\\
& &\frac{D n}{D h}  = 
- \frac{\partial^2\Omega_{S}}{\partial\mu\partial h} +
\frac{\partial^2\Omega_{S}}{\partial\mu\partial\Delta} \frac{\partial^2\Omega_{S}}{\partial\Delta\partial h } /\frac{\partial^2
  \Omega_{S}}{\partial\Delta^2}=\frac{D\delta n}{D \mu} \,,\\
& &\frac{D\delta n}{D h} =
- \frac{\partial^2\Omega_{S}}{\partial h^2} +
\left( \frac{\partial^2\Omega_{S}}{\partial\Delta\partial h }\right)^2 /\frac{\partial^2
  \Omega_{S}}{\partial\Delta^2} \,.
\end{eqnarray}
\end{subequations}

The expressions of the related second-order derivatives are 
\begin{eqnarray}
& &\frac{\partial^{2}\Omega_{S}}{\partial\Delta^{2}}=\sum_{\mathbf{k}}\Big\{\frac{\Delta^{2}}{E^{-2}_{k}}\big[\frac{1-f^{-}}{E^{-}_{k}}+f^{\prime -}\big]+\frac{\Delta^{2}}{E^{+2}_{k}}\big[\frac{1-f^{+}}{E^{+}_{k}}+f^{\prime +}\big]\Big\},\nonumber \\
& &\frac{\partial^{2}\Omega_{S}}{\partial\mu^{2}}=-\sum_{\mathbf{k}}\Big[\frac{\Delta^{2}}{E^{-3}_{k}}\big(1-f^{-}\big)-\frac{\xi^{-2}_{ka}}{E^{-2}_{k}}f^{\prime -}+\frac{\Delta^{2}}{E^{+3}_{k}}\big(1-f^{+}\big)-\frac{\xi^{+2}_{ka}}{E^{+2}_{k}}f^{\prime +}\Big],\nonumber \\
& &\frac{\partial^{2}\Omega_{S}}{\partial h^{2}}=\sum_{\mathbf{k}}\big[f^{\prime-}+f^{\prime+}\big],\nonumber \\
& &\frac{\partial^{2}\Omega_{S}}{\partial\Delta\partial\mu}=-\sum_{\mathbf{k}}\Big\{\frac{\xi^{-}_{ka}\Delta}{E^{-2}_{k}}\big[\frac{1-f^{-}}{E^{-}_{k}}+f^{\prime-}\big]+\frac{\xi^{+}_{ka}\Delta}{E^{+2}_{k}}\big[\frac{1-f^{+}}{E^{+}_{k}}+f^{\prime+}\big]\Big\},\nonumber \\
& &\frac{\partial^{2}\Omega_{S}}{\partial\Delta\partial h}=-\Delta\sum_{\mathbf{k}}\Big[\frac{\bar{f}^{\prime-}}{E^{-}_{k}}-\frac{\bar{f}^{\prime+}}{E^{+}_{k}}\Big], \nonumber\\
& &\frac{\partial^{2}\Omega_{S}}{\partial\mu\partial h}=\sum_{\mathbf{k}}\Big[\frac{\xi^{-}_{ka}}{E^{-}_{k}}\bar{f}^{\prime-}+\frac{\xi^{+}_{ka}}{E^{+}_{k}}\bar{f}^{\prime+}\Big].
\end{eqnarray}
Here $f^{+}\equiv f(E^{+}_{k\alpha})+f(E^{+}_{k\beta})$, $f^{-}\equiv f(E^{-}_{k\alpha})+f(E^{-}_{k\beta})$, $f^{\prime +}\equiv f^{\prime}(E^{+}_{k\alpha})+f^{\prime}(E^{+}_{k\beta})$, $f^{\prime -}\equiv f^{\prime}(E^{-}_{k\alpha})+f^{\prime}(E^{-}_{k\beta})$, $\bar{f}^{\prime+}\equiv f^{\prime}(E^{+}_{k\alpha})-f^{\prime}(E^{+}_{k\beta})$, $\bar{f}^{\prime-}\equiv f^{\prime}(E^{-}_{k\alpha})-f^{\prime}(E^{-}_{k\beta})$, and $f^{\prime}(x)=df(x)/dx$. 

\bibliographystyle{apsrev}

\end{document}